\magnification=1200
\input iopppt.modifie
\input epsf
\def\received#1{\insertspace 
     \parindent=\secindent\ifppt\textfonts\else\smallfonts\fi 
     \hang{#1}\rm } 
\def\appendix{\goodbreak\beforesecspace 
     \noindent\textfonts{\bf Appendix}\secspace} 
\def\figure#1{\global\advance\figno by 1\gdef\labeltype{\figlabel}%
   {\parindent=\secindent\smallfonts\hang 
    {\bf Figure \ifappendix\applett\fi\the\figno.} \rm #1\par}} 
\headline={\ifodd\pageno{\ifnum\pageno=\firstpage\titlehead
   \else\rrhead\fi}\else\lrhead\fi}

\def\rrhead{\textfonts\hskip\secindent\it 
    \shorttitle\hfill\rm L\folio} 
\def\lrhead{\textfonts\hbox to\secindent{\rm L\folio\hss}%
    \it\aunames\hss} 
\footline={\ifnum\pageno=\firstpage
\hfil\textfonts\rm L\folio\fi}   
\def\titlehead{\smallfonts J. Phys. A: Math. Gen. {\bf 25} (1992)
L127--L134  \hfil} 

\firstpage=127
\pageno=127

\jnlstyle
\jl{1}
\overfullrule=0pt

\letter{Anisotropic critical phenomena in parabolic 
geometries: the directed self--avoiding walk}[Letter to
the Editor]

\author{Lo\"\ii c Turban}[Letter to the Editor]
 
\address{Laboratoire de Physique du Solide\footnote{\dag}{Unit\'e de
Recherche associ\'ee au CNRS no 155}, Universit\'e de Nancy I, BP 239
\hfil\break F--54506 Vand\oe uvre l\'es Nancy Cedex, France}

\received{Received 30 September 1991}

\abs
The critical behaviour of directed self--avoiding walks is studied 
on parabolic--like systems with a free boundary at $x\!=\!\pm
Ct^\alpha$.  Using a scaling argument, $1/C$ is shown to be a
marginal variable  when $\alpha\!=\!\nu_\perp/\nu_\parallel\!=\!1/2$
i.e. on a parabola. As  a consequence the directed walk may display
varying local exponents. Such a behaviour is indeed observed for
restricted walks.  This generalizes a result of Cardy showing that
nonuniversal  behaviour occurs at corners for isotropic systems.
\endabs

\vglue1cm


\submitted
\date

Isotropic systems are known to display nonuniversal critical behaviour
at corners [1--6], local exponents varying 
continuously with the opening angle $\theta$. As shown recently [7], 
this may be linked to the scale invariance of these shapes in the 
case of isotropic systems. In parabolic--like geometries where the 
boundary is located at $X(t)\!=\!\pm Ct^\alpha $
an isotropic change of scale transforms $C$ into $b^{\alpha-1}C$ 
where $b$ is the dilatation factor. The dimension of $C$ vanishes 
when $\alpha\!=\!1$, i.e. in the corner geometry, then 
$C\!=\!\tan\theta/2$ is a marginal variable leading to 
$\theta$-dependent exponents. When $\alpha\!>\!1$, $C$ grows under 
renormalization and the critical behaviour is that of a flat 
surface whereas when $\alpha\!<\!1$, $C$ decreases and one gets 
either a line geometry or a cut, depending on the location of 
the system relative to its boundary.

These considerations may be extended to the case of anisotropic 
systems [8] for which the correlation length diverges at the critical 
point with different critical exponents in the parallel and 
perpendicular directions. If the $t$-axis of the boundary defined 
above coincides with the parallel direction of the system with a 
correlation length exponent $\nu_\parallel\!=\!z\nu_\perp$, under 
an anisotropic change of scale [9] with dilatation factors 
$b_\parallel\!=\!b^z$ and $b_\perp\!=\! b$, $C$ is changed into 
$$
C'=b^{z\alpha -1}C \eqno(1)
$$
and marginal behaviour is obtained when $\alpha\!=\!1/z$.

In the present work, we check these ideas in the case of the directed
self--avoiding walk [8,10] for which $z\!=\!2$ so that varying
exponents are expected inside a parabola. Due to the directedness,
one cannot get any boundary effect for a walk outside a parabola.
Although in  the following we restrict ourselves to the
{\smallfonts2D} problem, similar  results are expected in higher
dimensions.

Let us consider a directed walk on a rectangular lattice in the 
$(x,t)$ plane. At each step with $\Delta t\!=\!\tau$ in the time 
direction, the walker performs a jump $\Delta x\!=\!\pm a$ towards 
one of the two nearest sites with the same probability so that the 
walk is restricted and may be also considered as a {\smallfonts1D} diffusion 
process. Furthemore the walker is assumed to start at the origin
$(x\!=\! t\!=\!0)$ and remains confined inside a parabolic--like
domain which for convenience we consider to be slightly shifted
backwards in  time 
$$
\vert x(t)\vert <X(t)=C(t+\eta )^\alpha \eqno(2)
$$
The number of $N$-step walks starting from the origin and reaching 
$x\!=\! na$ at time $t\!=\! N\tau$ may be written as
$$
{\cal N}_N(0,na)=2^NP(na,N\tau)\eqno(3)
$$
where the front factor on the right gives the total number of 
unconfined walks with $N$ steps and $P(x,t)$ gives the 
probability to reach $x=na$ without crossing the frontier. This
probability satisfies the recursion equation
$$
P(x,t+\tau)={1\over 2}\left[P(x+a,t)+P(x-a,t)\right]\qquad 
P(x,0)=\delta_{x,0}\eqno(4)
$$
with the boundary condition 
$$
P(x,t)=0\qquad \vert x\vert = X(t)\eqno(5)
$$
In the continuum limit $(a\!\ll\!1$, $\tau\!\ll\!1$,
$a^2/\tau\!=\!1)$, one gets the diffusion equation 
$$
{\partial P\over \partial t}={1\over2}{\partial^2P\over \partial
x^2}\eqno(6)
$$
together with the time--dependent absorbing boundary condition given 
in (5).

In a free random walk $\overline{x^2(t)}\!=\! t$ so that when 
$\alpha\!>\!1/2$ the walker no longer sees the frontier when
$t\!\gg\!t^*\!=\! C^{2/(1\!-\!2\alpha )}$ and the probability
distribution tends asymptotically to the Gaussian form, namely
$$
P(x,t)\sim {1\over \sqrt{ 2\pi t}}\exp\left(-{x^2\over 2t}\right)
\qquad t\gg t^*\eqno(7)
$$
This is the behaviour expected from equation (1) since $C$ grows 
under renormalization when $\alpha\!>\!1/2$ so that the boundary 
evolves towards a flat surface geometry.

When $\alpha\!\leq\!1/2$ the boundary changes the asymptotic behaviour
and in order to work with constant boundary conditions, it is convenient
to introduce the new variable 
$$
y={x\over (t+\eta )^\alpha}\eqno(8)
$$
leading to the following equation for $P(y,t)$
$$
(t+\eta)^{2\alpha}{\partial P\over \partial t}=
{1\over 2}{\partial^2P\over \partial y^2}+\alpha y(t+\eta )^
{2\alpha -1}{\partial P\over \partial y}\eqno(9)
$$
with absorbing boundary conditions at $y\!=\!\pm C$.

In the appendix, scaling arguments are used to get the form of the 
probability distribution. Its leading  behaviour at long 
time when $\alpha\!<\!1/2$ is found to take the form
$$
P(x,t)\simeq {1\over Ct^\alpha}\exp \left(-{\pi^2\over 8C^2}
{t^{1-2\alpha }\over 1-2\alpha }\right)\cos\left({\pi x\over 
2Ct^\alpha}\right)\eqno(10)
$$
This asymptotic expression becomes exact in the limit of a 
strip geometry when $\alpha\!=\!0$. A similar stretched exponential 
behaviour was previously observed in isotropic parabolic systems 
with relevant boundary effects [7].

In the marginal case $\alpha\!=\!1/2$ the variables separate in 
(9) and the problem is exactly solvable. The diffusion 
equation becomes
$$
(t+\eta){\partial P\over \partial t}={1\over 2}
{\partial^2P\over \partial y^2}+{y\over 2}{\partial P\over 
\partial y}\eqno(11)
$$
and looking for $P$ as a product $\phi(t) \psi(y)$ enables us to
write down
$$
(t+\eta){\d\phi\over \d t} =-\lambda^2\phi\eqno(12{\rm a})
$$
$$
z{\d^2\psi \over \d z^2}+\left({1\over 2}-z\right){\d \psi\over
\d z}-\lambda^2\psi =0\eqno(12{\rm b})
$$
where we used the new variable $z\!=\!-y^2/2$ in the eigenvalue
equation  (12{\rm b}). One recognizes Kummer's equation [11] so that
the eigenfunctions are confluent hypergeometric functions. 
Even solutions $_1F_1(\lambda^2,{1\over 2};z)$ must be 
selected since the initial condition introduces a reflection 
symmetry with respect to the time axis. These have the following 
z-expansion
$$
\fl _1F_1(a,b;z)=1+{a\over b}z+{a(a+1)\over b(b+1)}{z^2\over
2!}+\cdots\cr
\qquad\qquad+{a(a+1)(a+2)\cdots (a+n-1)\over b(b+1)(b+2)\cdots
(b+n-1)} {z^n\over n!}+\cdots\eqno(13)
$$
In order to satisfy the boundary conditions, the eigenvalues 
$\lambda_m^2$ have to belong to a discrete set which, according 
to (2) and (8), corresponds to zeros of the confluent hypergeometric 
function for $z\!=\!-C^2/2$. They are solutions of the implicit
equation 
$$
_1F_1\left(\lambda_m^2,{1\over 2};-{C^2\over 2}\right)=0\eqno(14)
$$
The first differential equation (12{\rm a}) simply yields 
$$
\phi_m(t)\sim (t+\eta )^{-\lambda_m^2}\eqno(15)
$$
and putting these together, one may finally write
$$
P(x,t)=\sum_mA_m(\eta)(t+\eta)^{-\lambda_m^2}{_1}F_1
\left(\lambda_m^2,{1\over 2};-{x^2\over 2(t+\eta )}\right)\eqno(16)
$$
where the coefficients $A_m(\eta )$ have to be chosen to satisfy
the  initial condition. The asymptotic behaviour is governed by the
lowest eigenvalue $\lambda_0^2$ which was studied numerically
(figure 1). 
{\par\begingroup\parindent=0pt\medskip
\epsfxsize=9truecm
\topinsert
\centerline{\epsfbox{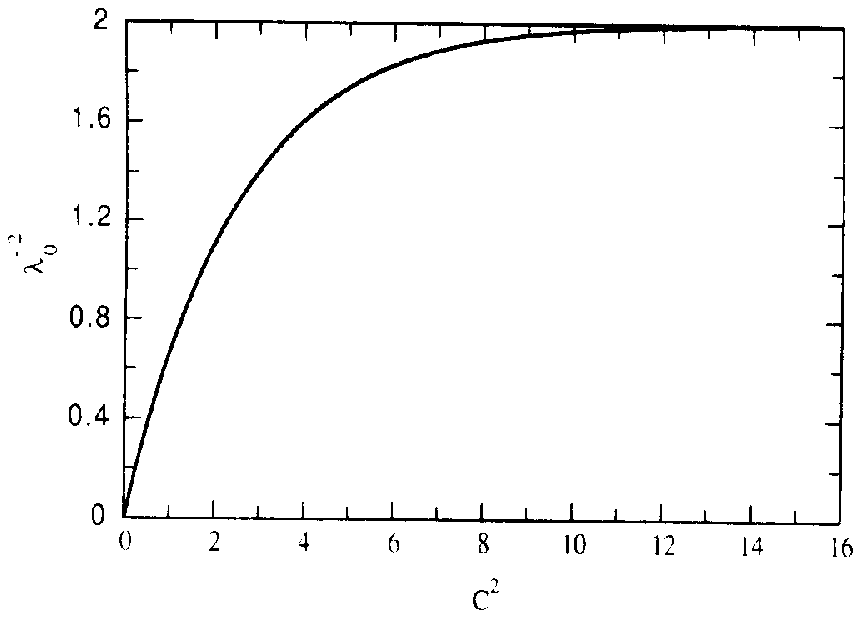}}
\smallskip
\figure{Inverse of the largest eigenvalue of equation (12b) as a 
function of the square of the marginal parameter $C$.}   
\endinsert 
\endgroup
\par}
Some analytical results may be obtained only in limiting cases.
Let us first consider the strong curvature limit
$\varepsilon\!=\! C^2/2\!\to\!0$. Then, according to (13) and
(14) $\lambda_m^2$ is  $O(\varepsilon^{-1})$ and introducing
$u\!=\! C^2\lambda_m^2=O(1)$ in  (14), one gets $\cos(\sqrt
{2u})\!=\!O(\varepsilon)$ so that the  eigenvalue spectrum reads
$$
\lambda_m^2=\left(m+{1\over 2}\right)^2{\pi^2\over 2C^2}+O(1)
\qquad m=0,1,2\cdots\eqno(17)
$$
The eigenfunctions are obtained in the same way leading to
$$
_1F_1\left(\lambda_m^2,{1\over 2};-{x^2\over 2(t+\eta )}\right)
=\cos\left[\left(m+{1\over 2}\right){\pi x\over C\sqrt{t+\eta}}
\right]+O(C^2)\eqno(18)
$$
The completness relation 
$$
{1\over C\sqrt \eta }\sum_{m=0}^\infty\cos\left[\left(m+{1\over 2}
\right){\pi x\over C\sqrt\eta}\right]=\delta (x)\eqno(19)
$$
may be used to deduce the coefficients $A_m(\eta)$ satisfying the
initial condition $P(x,0)\!=\!\delta (x)$ and the probability 
distribution behaves as
$$
\fl P(x,t)={1\over C\sqrt\eta }\sum_{m=0}^\infty \left(1+{t\over\eta}
\right)^{-\left(m+{1\over 2}\right)^2{\pi ^2\over 2C^2}}
\cos \left[\left(m+{1\over 2}\right){\pi x\over C\sqrt{t+\eta }}
\right]+O(C)\eqno(20)
$$
the asymptotic behaviour being governed by the first term in the
eigenvalue expansion, namely
$$
P(x,t)\sim t^{-{\pi^2\over 8C^2}}\cos\left({\pi x\over 2C\sqrt t}
\right)\qquad t\gg 1\eqno(21)
$$
This is just the result given by (10) when $\alpha\!\to\!1/2$
if $t^{1-2\alpha}$ is changed into $t^{1\!-\!2\alpha}\!-\!1$  in the
exponential, which amounts to modify a constant prefactor,  in order
to get a meaningfull limit.

Let us now consider the opposite limit. First, when C is infinite,
the boundary condition is satisfied with $\lambda_0^2\!=\!1/2$ since
[11] 
$$
_1F_1\left({1\over 2},{1\over 2};z\right)=\e^z\eqno(22)
$$
which corresponds to the Gaussian distribution
$$
P(x,t)={1\over \sqrt{2\pi (t+\eta)}}\exp\left[-{x^2\over 2(t+\eta)}
\right]\eqno(23)
$$
satisfying the initial condition when the limit $\eta\!\to\!
0$ is taken. This is the free diffusion result (7) as expected for a 
walker starting on a surface which is flat in this limit. When the 
curvature is weak $(C\!\gg\!1)$ one may use the $z^{-1}$-expansion of 
the confluent hypergeometric function
$$
\fl _1F_1(a,b;z)={\Gamma (b)\over \Gamma (a)}e^zz^{a-b}[1+O(\vert
z\vert ^{-1})]+{\Gamma (b)\over \Gamma (b-a)}(-z)^{-a}[1+O(\vert
z\vert  ^{-1})]\eqno(24)
$$
to find out the leading correction to $1/2$. After some algebra one 
gets
$$
\lambda_0^2={1\over 2}+{C\over \sqrt{2\pi }}\exp\left(-{C^2\over 2}
\right)\left[1+O(\varepsilon )\right]\eqno(25)
$$
where $\varepsilon$ is the correction term itself.

Let us finally turn to the evaluation of the critical exponents. 
Besides the radius of gyration exponents $\nu_\parallel\!=\!1$ and 
$\nu_\perp\!=\!1/2$, in analogy with thermal critical phenomena,
one defines a bulk susceptibility exponent $\gamma$ which enters
into the asymptotic behaviour of the total number of $N$-step 
directed walks starting from the origin on an unlimited lattice,  
${\cal N}_N\!\sim\!\mu^NN^{\gamma\!-\!1}$ [8]. Since for the
restricted  walk, $N=t$ when the time is measured in $\tau$ units,
by (3) this  can be written as
$$
{\cal N}_N=2^N\int_{-\infty }^{+\infty}P(x,t)\d x=2^N\eqno(26)
$$
so that $\mu\!=\!2$ and $\gamma\!=\!1$. Surface exponents may be
defined  in a system with a straight free surface along the time axis
by considering the total number of $N$-step walks starting near
the surface at $x\!=\!\delta $
$$
\int_0^\infty{\cal N}_N(\delta ,x)\d x\sim\mu^NN^{\gamma_1-1}\eqno(27)
$$
and the number of $N$-step walks starting and ending near the surface
$$
{\cal N}_N(\delta ,\delta )\sim\mu^NN^{\gamma_{11}-1}\eqno(28)
$$
Using the Gaussian distribution (7) and the method of images [10], 
it is easy to check that $\gamma_1\!=\!1/2$ and
$\gamma_{11}\!=\!-1/2$ in agreement with the scaling law
$2\gamma_1\!-\!\gamma_{11}\!=\!\gamma\!+\!\nu_\perp$ [8]. 
Now, in the parabolic geometry, one may define 
two new exponents by considering either the total number of 
$N$-step walks starting near the tip 
$$
\int_{-Ct^\alpha}^{+Ct^\alpha}{\cal N}_N(0,x)\d x\sim\mu^NN^
{\gamma_0-1}\eqno(29)
$$
or the number of $N$-step walks starting near the tip and ending
near the boundary
$$
{\cal N}_N(0,Ct^\alpha -\delta )\sim\mu^NN^{\gamma_{01}-1}\eqno(30)
$$
Using the analogy with thermal critical phenomena, it may be shown
that when $1/2\!\geq\!\alpha\!>\!0$ these exponents satisfy the
scaling law  
$$
\gamma_0-\gamma_{01}=\gamma_1-\gamma_{11}\eqno(31)
$$
since the boundary is asymptotically flat and parallel to the time 
axis.
When $\alpha\!>\!1/2$ one evidently gets $\gamma_0\!=\!\gamma\!=\!1$
and  $\gamma_{01}$ remains undefined since at long time the walks
cannot reach the surface.
In the marginal case, $\alpha\!=\!1/2$, the leading contribution 
to $P(x,t)$ in (16) gives
$$
\int_{-C\sqrt t}^{+C\sqrt t}{_1}F_1\left(\lambda_0^2,{1\over 2};-{x^2
\over 2t}\right)\d x\sim t^{1/2}\eqno(32)
$$
which combines with (3), (16) and (29) to yield
$$
\gamma_0={3\over 2}-\lambda_0^2\eqno(33)
$$
The second exponent is obtained in the same way through a
first--order expansion of the confluent hypergeometric function.
Then, by (14) 
$$
_1F_1\left(\lambda_0^2,{1\over 2};-{C^2\over 2}+{C\delta\over\sqrt t}
\right)\sim t^{-1/2}\eqno(36)
$$
and finally
$$
\gamma_{01}={1\over 2}-\lambda_0^2\eqno(35)
$$
follows in agreement with the scaling law (31). 
When $\alpha\!<\!1/2$, by (3) and (10), the connective constant $\mu$
is  given by
$$
\ln\mu=\ln 2-\lim_{N\to\infty}{\pi^2\over 8C^2}{N^{-2
\alpha }\over 1-2\alpha }\eqno(36)
$$
so that $\mu\!=\!2$ when $\alpha\!>\!0$ and due to the exponential
decay of  the probability distribution,
$\gamma_0\!=\!\gamma_{01}\!=\!-\infty$. It  follows that when C varies
from zero to infinity, the marginal  exponents interpolate between
their values below and above $\alpha\!=\!1/2$.
When $\alpha\!=\!0$, by (36), the connective constant is changed into
$$
\mu =2\exp \left(-{\pi^2\over 8C^2}\right)\eqno(37)
$$
and as a consequence of the one--dimensionality of the system 
$\gamma_0\!=\!\gamma_{01}\!=\!1$. Let us mention that doing the
calculation  with the transfer matrix technique on a lattice would
give $\mu\!=\!2\cos(\pi/2(C\!+\!1))$ instead of (37) but
the  continuum limit used here is valid for $a\!\ll\!1$ so that both 
expressions should be compared in the limit $C\!\gg\!1$ where they 
indeed give the same result. Finally, when $\alpha\!<\!0$, the
systems  shrinks at long time and according to (36) the connective
constant  vanishes. 

\ack
The author enjoyed long discussions with Ingo Peschel and Ferenc 
Igl\'oi about critical phenomena in parabolic geometries. 

\appendix

As suggested in the introduction the shape of the system may
be considered as a perturbation to its critical behaviour in infinite 
geometry characterized by the scaling field $1/C$ which, 
according to (1), may be either relevant, marginal or irrelevant
depending on the sign of its scaling dimension $1\!-\! z\alpha$. 
This allows us to write down a scaling ansatz for the probability
distribution
$$
P\left(x,t,{1\over C}\right)=b^{-1}P\left({x\over b},{t\over b^z},
{b^{1-z\alpha}\over C}\right)\eqno({\rm A1})
$$
where the scaling dimensions are those of the unperturbed fixed 
point. 

With $b\!\!=Ct^\alpha $ and $z\!=\!2$ for the directed walk, (A1)
translates  into 
$$
P\left(x,t,{1\over C}\right)={1\over Ct^\alpha}f\left({x\over Ct^
\alpha},{t^{1-2\alpha}\over C^2}\right)\eqno({\rm A2})
$$
so that for an irrelevant perturbation, a crossover towards the 
unperturbed critical behaviour occurs at $t^*\!=\!C^{2/(1-2\alpha)
}$  and a comparison with (7) shows that, in this case, the scaling 
function behaves asymptotically as a Gaussian
$$
f(u,v)={1\over \sqrt{2\pi v}}\exp\left(-{u^2\over 2v}\right)\eqno
({\rm A3})
$$
Furthermore (A2) can be inserted into (9) to get the asymptotic 
behaviour of the probability distribution in the case of a relevant 
perturbation. The scaling function then satisfies
$$
(1-2\alpha)v{\partial f\over \partial v}-\alpha f={1\over 2}v
{\partial^2f\over\partial u^2}+\alpha u {\partial f\over\partial u}
\eqno({\rm A4})
$$
When $\alpha\!<\!1/2$, $v$ grows in time and the leading behaviour 
is obtained by keeping the two first terms on both sides to get
$$
(1-2\alpha)v{\partial f\over \partial v}={1\over 2}v
{\partial^2f\over\partial u^2}=-\lambda^2\eqno({\rm A5})
$$
which becomes exact in the strip geometry when $\alpha\!=\!0$. With
$f\!\sim\!\psi(u)\phi(v)$, $\psi$ even in $u$ and vanishing for
$u\!=\!1$,  we find
$$
f(u,v)=\sum_{m=0}^\infty\exp\left(-{\lambda_m^2v\over 1-2\alpha }
\right)\cos(\sqrt 2\lambda_mu)\eqno({\rm A6a})
$$
$$
\lambda_m=\left(m+{1\over 2}\right){\pi\over\sqrt 2}\qquad m=0,1,2
\cdots\eqno({\rm A6b})
$$
Together with (A2) this gives a properly normalized exact 
expression when $\alpha\!=\!0$ whereas the term $m\!=\!0$ in the 
eigenvalue expansion provides the leading contribution (10) to the 
probability distribution.

\references
\numrefjl{[1]}{Cardy J L 1983}{\JPA}{16}{3617}
\numrefjl{[2]}{Barber M N, Peschel I and Pearce PA 
1984}{J. Stat. Phys.}{37}{497}
\numrefjl{[3]}{Cardy J L 1984}{\NP\ \rm B}{240}{514}
\numrefjl{[4]}{Bariev R Z 1986}{Teor. Mat. Fiz.}{69}{149}
\numrefjl{[5]}{Kaiser C and Peschel I 1989}{J. Stat. Phys.}{54}{567}
\numrefjl{[6]}{Davies B and Peschel I 1991}{\JPA}{24}{1293}
\numrefjl{[7]}{Peschel I, Turban L and Igl\'oi F
1991}{\JPA}{24}{L1229} 
\numrefbk{[8]}{Privman V and \v Svraki\'c N M
1989}{Directed Models  of Polymers, Interfaces and Clusters (Lecture
Notes in Physics  vol 338)}{(Berlin: Springer)}
\numrefjl{[9]}{Binder K and Wang J S 1989}{J. Stat. Phys.}{55}{87}
\numrefjl{[10]}{Fisher M E 1984}{J. Stat. Phys.}{34}{667}
\numrefbk{[11]}{Abramowitz M and Stegun I A 1965}{Handbook of 
Mathematical Functions}{(New York: Dover) p 504}

\vfill\eject
\bye